\documentclass[12pt]{article}
\usepackage{graphicx}
\begin{document}

%--------------------------------------------------------------------

\begin{center}
{\large\bf QCD's Partner needed for Mass Spectra and Parton
       Structure Functions} \\
\vspace{5mm}
Y. S. Kim \\
Center for Fundamental Physics, University of Maryland,
College Park, Maryland 20742, U.S.A., yskim@umd.edu
\end{center}

\vspace{5mm}

\begin{abstract}

  As in the case of the hydrogen atom, bound-state wave functions
  are needed to generate hadronic spectra.  For this purpose,
  in 1971, Feynman and his students wrote down a Lorentz-invariant
  harmonic oscillator equation. This differential equation has
  one set of solutions satisfying the Lorentz-covariant boundary
  condition.  This covariant set generates Lorentz-invariant
  mass spectra with their degeneracies.  Furthermore, the
  Lorentz-covariant wave functions allow  us to calculate the
  valence parton distribution by Lorentz-boosting the quark-model
  wave function from the hadronic rest frame.  However, this
  boosted wave function does not give an accurate parton
  distribution.  The wave function needs QCD corrections to make
  a contact with the real world.  Likewise QCD needs the wave
  function as a starting point for calculating the parton
  structure function.
\end{abstract}
\vspace{16mm}
Presented at the Excited QCD, Zakopane, Poland (February 2009).

\newpage
At the 1965 meeting of the American Physical Society held in
Washington, DC, U.S.A.  Freeman Dyson stated that quantum
electrodynamics can become more effective if combined with other
theories~\cite{dyson65}.  He was right, but he gave a wrong
example.  He mentioned the calculation of the neutron-proton mass
difference by Dahsen and Frautchi as an example.  It is still
believed that the neutron and proton have the same mass, and
the mass difference comes from an electromagnetic perturbation.
However, their perturbation calculation uses a non-localized
a wave function which increases exponentially for large
distance~\cite{kim66}.

Dyson was still right in saying that QED needs a partner to
be most effective.  The partner is the localized bound-state
wave function.  Let us look at the Lamb shift.  QED gives to
the Coulomb potential a delta function correction at the origin.
The S state gets affected by this potential, while the P state
is insensitive to this correction at the origin.  This results
in the shifts between and P and S states.  The Lamb shift is
regarded as one of the triumphs of quantum electrodynamics.

Indeed, in order to calculate the Lamb shift, we need hydrogen
wave functions, but quantum electrodynamics cannot produce
localized wave functions with probability interpretation.  We
still have to solve the wave equation with the standing-wave
boundary condition to get the Rydberg energy levels and
corresponding wave functions.

QED with Feynman diagrams is designed to address scattering
problems in the Lorentz-covariant world.  The situation is the
same in QCD, which is an extension of QED with gluon instead
of photons.  QCD can make corrections to the existing mass
spectra and structure functions, but cannot produce wave
functions with proper boundary conditions.  Thus, QCD alone
cannot produce hadronic mass spectra or parton distributions.
It needs a partner.

%---------------------------------------------------------------
\begin{figure}%[thb]
\centerline{\includegraphics[scale=0.5]{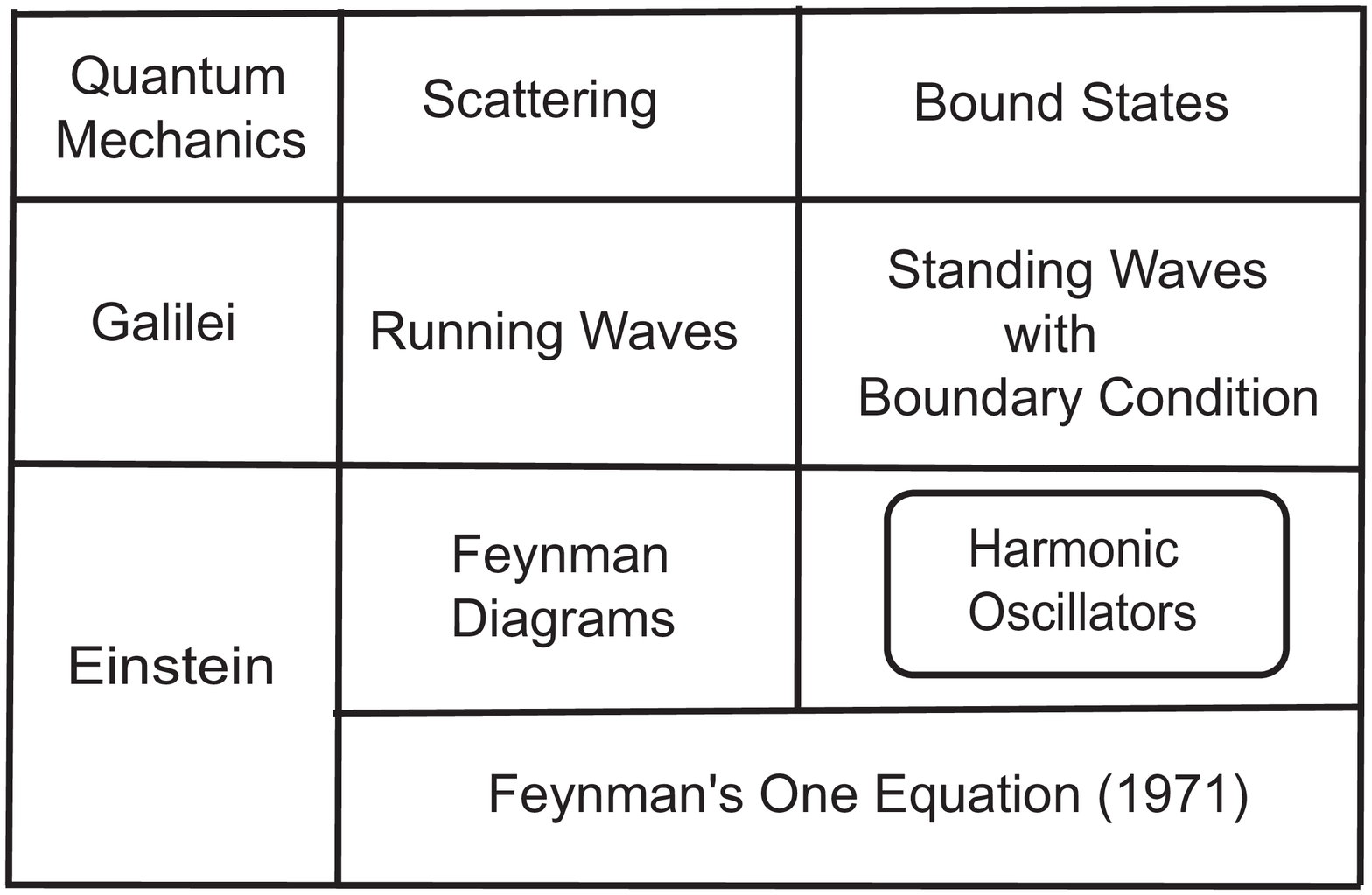}}
\caption{Quantum mechanics in Galilei and Einstein systems.
It is possible to construct a Lorentz-covariant model of bound
states.  Feynman and his students in 1971 wrote down a
Lorentz-invariant differential equation which contains both
running and standing waves.}\label{comet77}
\end{figure}
%-----------------------------------------------------------------

In 1971, Feynman and his students noted that harmonic oscillator
wave functions with their three-dimensional degeneracy can explain
the main features of the hadronic spectra~\cite{fkr71}.  Earlier
in 1969~\cite{fey69}, Feynman proposed his parton picture where
a fast-moving hadrons appears like a collection of partons with
properties quite different from those of the quarks inside a
static hadron.

In their 1971 paper~\cite{fkr71}, Feynman~{\it et al.} wrote down
the Lorentz-invariant equation which can be separated into the
Klein-Gordon equation for a free hadron, and a harmonic-oscillator
equation for the quarks inside the hadron, which determines the
hadronic mass.  Feynman's equation of 1971 contains both running
waves for the hadron and the standing waves for the quarks inside
the hadron, as indicated in Fig.~\ref{comet77}.

The oscillator equation takes the form
\begin{equation}\label{fkr11}
\frac{1}{2} \left[\left(\frac{\partial}{\partial x_{\mu}}\right)^2
- x_{\mu}^2 \right] \psi\left(x_{\mu}\right) =
\lambda \psi \left(x_{\mu}\right) ,
\end{equation}
where $x_{\mu}$ is the four-vector specifying the space-time
separation between the quarks.  For convenience, we ignore all
physical constants such as $c, \hbar$, as well as the spring
constant for the oscillator system.

In the hadronic rest frame, if the time-like excitations are
suppressed, this equation produces hadronic mass spectra~\cite{fkr71}.
If the hadron starts moving along the $z$ direction, we can separate
out the transverse coordinates $x$ and $y$, and write the differential
equation of Eq.(\ref{fkr11}) as
\begin{equation}\label{fkr22}
\frac{1}{2} \left[-\left(\frac{\partial}{\partial z}\right)^2 + z^2
 +\left(\frac{\partial}{\partial t}\right)^2 - t^2 \right]
\psi(z,t) = \lambda \psi(z,t) ,
\end{equation}
where $t$ is the time-separation variable between the quarks.  From
this equation, Feynman~{\it et al.} wrote down their solution
\begin{equation}\label{fkr33}
\psi(z,t) = \exp{\left\{-\frac{1}{2}\left(z^2 - t^2\right)\right\}}.
\end{equation}
This form is a Gussied function for the space-like $z$ coordinate if
the time-like variable $t$ is ignored.  It is also invariant under
Lorentz boosts along the $z$ direction.  However, due to its non-local
time-like distribution, this expression cannot be regarded as a
physically meaningful wave function.

On the other hand, this equation also has a solution of the form
\begin{equation}\label{kn11}
\psi(z,t) = \exp{\left\{-\frac{1}{2}\left(z^2 + t^2\right)\right\}}.
\end{equation}
This solution is Gaussian in both the $z$ and $t$ variables.  Is
it then possible to attach a physical interpretation to this wave
function.

\begin{figure}%[thb]
\centerline{\includegraphics[scale=0.5]{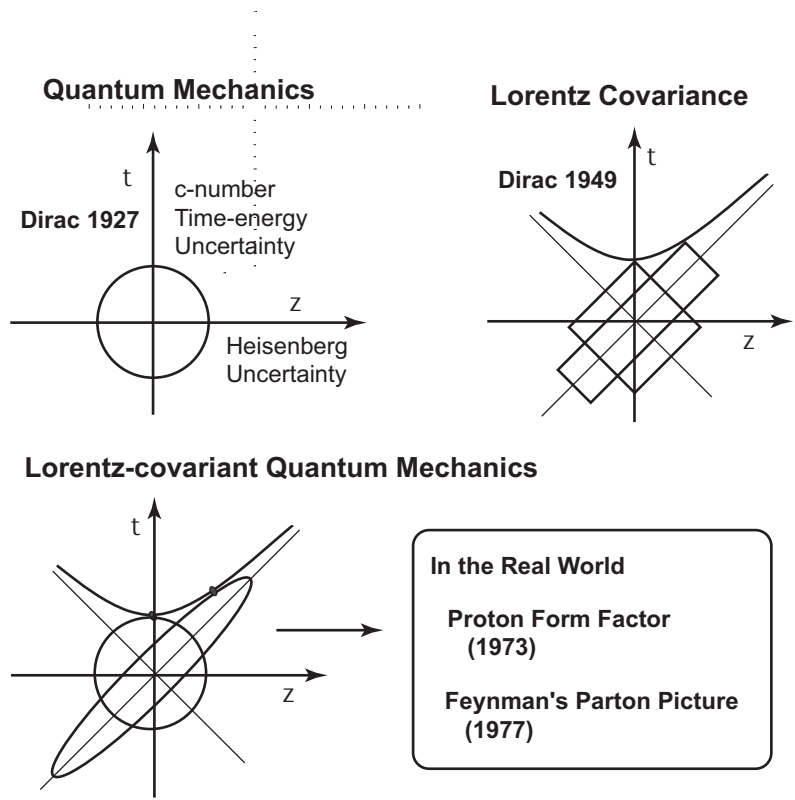}}
\vspace{5mm}
\caption{Space-time picture of quantum mechanics.  In his 1927,
Dirac noted that there is a c-number time-energy uncertainty
relation, in addition to Heisenberg's position-momentum uncertainty
relations, with quantum excitations.  This idea is illustrated
in the first figure.  In 1949, Dirac produced his light-cone
coordinate system as illustrated in the second figure.  It is
then not difficult to produce the third figure, for a
Lorentz-covariant picture of quantum mechanics.}\label{kn73}
\end{figure}

First, the time-separation $t$ exists wherever there is a space
separation, according to Einstein.  According to quantum mechanics,
there is a time-energy uncertainty relation associated with this
variable, as shown in Fig.~\ref{kn73}.

As Dirac noted in 1927~\cite{dir27}, this time-energy uncertainty
does not cause excitations, while Heisenberg's uncertainty generate
excitations along the space-like $z$ axis. However, this space-time
asymmetry is quite consistent the internal space-time symmetries
dictated by Wigner's little group~\cite{wig39,knp86}.  According to
Wigner, the internal space-time symmetry of massive particles is
that of the three-dimensional rotation group without the time
variable.  We can summarize these in terms of the circle given
in Fig.~\ref{kn73}.

How about the Lorentz invariance?  The form given in Eq.(\ref{fkr33})
is invariant under Lorentz boosts as $\left(z^2 - t^2\right)$ is.
However, the expression $\left(z^2 + t^2\right)$ in Eq.(\ref{kn11})
is not invariant.  Is this the end of the story?  No!  Let us
boost this form using Dirac's light-cone system~\cite{dir49}.

%---------------------------------------------------------------------
%---------------------------------------------------------------------

If the hadron moves along the $z$ direction with the velocity
parameter $\beta$, the wave function of Eq.(\ref{kn11}) becomes
\begin{equation}\label{kn22}
\exp{\left\{-\frac{1}{4}\left[\frac{1 - \beta}{1 + \beta}(z + t)^2
 + \frac{1 + \beta}{1 - \beta}(z - t)^2\right] \right\}},
\end{equation}
This is an elliptic distribution given in Fig.~\ref{kn73}, where
the circular distribution is modulated by Dirac's light-cone
picture of Lorentz boosts.  The circle is ``squeezed'' into the
ellipse.

The question is whether we can see the effects of this Lorentz
squeeze in the real world.  In 1973~\cite{kn73ff}, in terms of
Lorentz-squeezed hadrons,  Kim and Noz were able to explain the
form factor calculation of Fujimura, Kobayashi, and Namiki who
derived the dipole cut-off of the proton form factor for large
momentum transfers~\cite{fuji70}.

According to Fig.~\ref{kn73}, the quark distribution becomes
concentrated along the immediate neighborhood of one of the
light cones as the hadronic speed becomes closer to that of
light.  In 1977~\cite{knp86,kn77par}, Kim and Noz were able to
explain the peculiarities of Feynman's parton picture.  Partons
have the following peculiar properties.
\begin{itemize}

 \item[1.] Partons are like free particles, unlike the quarks
          inside a hadron.

 \item[2.] The parton distribution function becomes wide-spread
          as the hadron moves faster.  The width of the distribution
          is proportional to the hadron momentum.

 \item[3.] The number of partons appears to be infinite.

\end{itemize}
In the ellipse given in Fig.~\ref{kn73}, one of the axis becomes
longer while the other becomes shorter. In 2005~\cite{kn05job},
Kim and Noz were able to associate these axes as the interaction
time between the quarks and the interaction time of one of the
quarks with the external signal, respectively.  Thus, the external
signal is not able to sense other quarks in the hadron.  This is
what Feynman said in his original papers on the parton
model~\cite{fey69}.

Kim and Noz indeed explained all the peculiarities of Feynman's
parton picture, and proved that the quark model and the parton
model are two different manifestations of one Lorentz-covariant
entity.  However, is it possible to calculate the parton
distribution function by boosting the quark wave function from
the rest frame?  The hadron, when it moves fast, contains both
valence partons and gluonic partons.  We should therefore obtain
the valence parton distribution by boosting the rest-frame wave
function.

\begin{figure}%[thb]
\centerline{\includegraphics[scale=0.45]{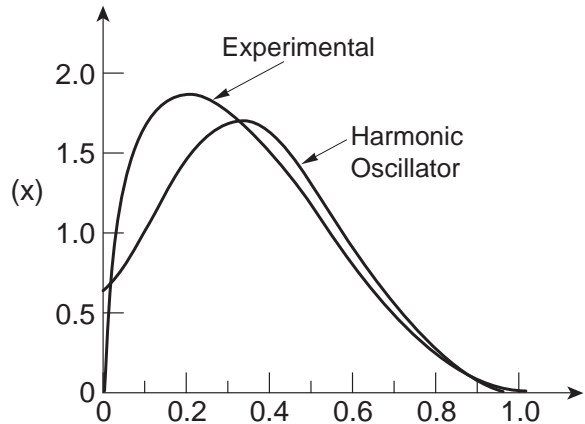}}
\vspace{5mm}
\caption{Parton distribution function from Hussar's
  paper~\cite{hussar81}.  Although there is a general agreement
  between theory and experiment, the disagreement is substantial.
  This difference  could be corrected by QCD.}\label{hussar}
\end{figure}

In 1980~\cite{hwa80}, Hwa observed that the external signals
do not directly interact with the quarks, but with dressed quarks
called valons.  Thus, if we remove the valon effect, we should
be able to measure the distribution of valence quarks. With this
point in mind, Hussar in 1981 compared the parton distribution
from the boosted oscillator wave function and the experimentally
measured distribution~\cite{hussar81}.  Hussar's result is
given in Fig.~\ref{hussar}.

As we can see in this figure, there is a general agreement
between the experimental data and the theoretical curve derived
from the static quark distribution.  Yet, the disagreement is
substantial, and this is the gap QCD has to feel in.  This
work is yet to be carried out.  The wave function needs QCD to
make contacts with the real world.  Likewise, QCD needs the
wave function as a starting point for calculating the parton
distribution.  They are need each other.  They are the
partners.

%--------------------------------------------------------------
%---------------------------------------------------------------

\end{document}